# Fast AC Steady-State Power Grid Simulation and Optimization Using Prior Knowledge

Aayushya Agarwal, Amritanshu Pandey, Larry Pileggi

*Abstract*— Fast and accurate optimization and simulation is widely becoming a necessity for large scale transmission resiliency and planning studies such as N-1 SCOPF, batch contingency solvers, and stochastic power flow. Current commercial tools, however, prioritize speed of convergence over accuracy by relying on initial conditions that are taken from the steady state solution of similar network configurations that are not guaranteed to lie within a convex region of a valid solution. In this paper we introduce a globally convergent algorithm to facilitate fast and accurate AC steady state simulation and optimization based on prior knowledge from similar networks. The approach uses a homotopy method that gradually and efficiently translates a previously known network configuration to the current network configuration. The proposed formulation is highly scalable, and its efficacy is demonstrated for resiliency study and optimization of large networks up to 70k buses.

## I. INTRODUCTION

Fast and accurate steady state simulation and optimization is essential to study the response against contingencies and disruptions to the AC electric power grid. To ensure steady-state accuracy, engineers use AC power flow and AC constrained optimal power flow (OPF) methodologies; however, the speed of AC constrained simulation and optimization is a critical bottleneck for many operations and planning applications. Operation engineers use tools such as security constrained optimal power flow (SCOPF) and batch contingency solvers in real time to determine the resiliency of the network. The real time constraint on the simulation and optimization makes large scale AC power flow and OPF simulation difficult to complete. For instance, ARPA-E's GO competition highlights the difficulties of solving SCOPF in real-time operation [1] primarily due to time constraints. While planning engineers don't have similar real-time constraints, they perform a large number of optimization and simulation runs to account for various scenarios. A fast simulation or optimization solver is essential to lower the overhead and enable the ability to study a broader number of scenarios.

The speed of simulation or optimization for a steady state are incumbent upon proper initial conditions within a convex region of the solution. In the absence of such initial conditions, the inherent numerical methods within these tools are prone to divergence, numerical oscillation, or slow convergence due to limiting methods [2]-[3]. Attaining proper initial conditions in practice, however, is often difficult. Real-time operations often use the previous state of a network as an initial condition for optimization and power flow simulation [1]. Occasionally, the previous state may differ due to topological or demand/supply changes. Planning engineers face similar challenges as they rely on previous networks for initial conditions. For example, in optimizing a future network over multiple time periods, planning engineers use the steady state of the previous time period as an initial condition for the current solution.

To attain the required runtime efficiency, planning and operation tools in practice sacrifice accuracy of simulation by using relaxation techniques such as DC network constraints [4][5] or linearized AC constraints [6]-[14]. The inaccuracy of using DC constraints is well understood and often leads to practically infeasible dispatches [6]. In contrast, other techniques have guaranteed robust convergence of the AC network by essentially disregarding initial conditions with the use of homotopy methods [15]-[17], thereby avoiding the use of prior network knowledge at the expense of runtime efficiency. The challenge of maintaining accuracy using non-linear AC constraints while achieving the acceptable runtime is well studied [6]-[9] with a wide range of methodologies to tackle this problem. Many previous works have developed efficient algorithms to devise better initial conditions based on a relaxed version of the network such as a SOCP relaxation [4], however, they have not demonstrated the capability to scale to large systems such as synthetic footprint of the Eastern Interconnection with 70k buses. Certainly, the need for a fast power flow and optimal power flow solver is apparent, as evidenced by prior work toward developing approximative methods to perform stochastic studies [9][10] as well as real-time operation studies [11]-[14]. Previous work [13][14] has also utilized deep reinforcement learning algorithms to learn the response of a network and are quickly able to find a dispatch for a given network condition. Homotopy methods have also been studied, with success on developing a post-transmission line contingency optimal power flow that gradually reduces the effect of the transmission line to drive to the optimal solution

A. A. Author is with Electrical Engineering Department, Carnegie Mellon University, Pittsburgh, PA 15213 USA (e-mail: aayushya@andrew.cmu.edu).
A. P. Author is with Electrical Engineering Department, Carnegie Mellon University, Pittsburgh, PA 15213 USA (e-mail: amritanp@andrew.cmu.edu).
L. P. Author is with Electrical Engineering Department, Carnegie Mellon University, Pittsburgh, PA 15213 USA (e-mail: pileggi@andrew.cmu.edu).

[18]. Presently, however, such methods have been limited to solving optimal power flow with outages.

To achieve timely convergence of non-linear AC power flow and optimization problems, we introduce a globally convergent methodology that uses the solution from a similar topological network as an initial condition. This prior knowledge, which is generally available in most practical scenarios, captures a known feasible point in the solution space. Starting with this known solution of the previous network, we utilize a novel homotopy method that we refer to as Network-Stepping, which gradually transforms to the current network, thereby tracing a path through the solution space. This technique uses the prior knowledge to avoid unnecessary numerical oscillations or divergence and achieves faster convergence to the solution. The proposed Network-Stepping method translates any discrete change in the network as current injections that are iteratively reduced within a homotopy framework to guarantee global convergence. The framework is useful for both steady-state power flow and various optimization methods.

The practicality of this methodology is first highlighted with its operational applications in solving real-time optimal power flow, a batch contingency power flow from an optimal dispatch with real-time constraints and a stochastic steady-state study. We implement this methodology in the SUGAR power flow formulation [15]. The efficacy of the Network-Stepping method is readily apparent when solving large optimization or power flow studies, such as simulating and optimizing the Eastern Interconnection, as it provides scalable, efficient and guaranteed convergence. The general approach of Network-Stepping is applicable for various power flow and optimization settings, which could support the goal of efficient and robust tools for operations and planning.

Section II highlights the numerical methods used to solve power flow and optimization problems that serve as a basis for the rest of the paper. Section III describes the Network-Stepping methodology with a provable analysis of guaranteed convergence. Section IV demonstrates its practical applications for large optimization and power flow studies.

## II. Background on Steady State Numerical Methods

Steady state simulation and optimization frameworks for power grids inherently solve a set of non-linear equations that represent the steady state of the network or a stationary point for the optimization problem. To solve the set of equations, frameworks utilize numerical methods, most common of which is Newton-Raphson (N-R). Solution methodologies for power flow and optimization overviewed to aid in introducing the Network-Stepping approach.

### A. Power Flow Solver

The set of non-linear equations that used in current-injection power flow, $F_{pf}$ (1) represents the steady-state current mismatch at the real and imaginary nodes. A state vector for the problem, $X_{pf}$, includes the real and imaginary voltages of the grid ($V_r$ and $V_i$), as well as generator reactive power ($Q_g$) (2).

$$F_{pf}(X_{pf}) = 0 \tag{1}$$

$$X_{pf} = \begin{bmatrix} V_r \\ V_i \\ Q_g \end{bmatrix} \tag{2}$$

*1) Feasibility through Slack Injections*

For a given dispatch, networks are often infeasible either due to supply-demand mismatch or PV instability [17], which results in divergence when using the N-R method to solve for the steady-state. To capture the exact infeasibility (represented as a mismatch in KCL), we place slack current injections, $I_{slack}$, at each node [17] to measure the current mismatch. Minimizing these slack injections ensures convergence regardless of the feasibility of the network; i.e. zero slack injection for a feasible network or a non-zero value that captures the infeasibility. Mathematically, the power flow analysis is supplemented with a vector of slack injections that is minimized:

$$\min \|I_{slack}\|^2 \\ s.t. \quad F_{pf}(X_{pf}) + I_{slack} = 0 \tag{3}$$

To solve for the infeasibilities, we can solve the optimization problem of (3) by forming a Lagrange (4) and represent the first order optimality conditions using (5)-(7). The final set of equations corresponding to identifying infeasibility is shown in (8), where the set of equations, $F_{Ipf}$, for the infeasibility power flow has a state vector, $X_{Ipf}$.

$$\mathcal{L} = \|I_{slack}\|^2 + \lambda^T(F_{pf}(X_{pf}) + I_{slack}) \tag{4}$$

$$\nabla_\lambda \mathcal{L} = F_{pf}(X_{pf}) + I_{slack} = 0 \tag{5}$$

$$\nabla_{I_{slack}} \mathcal{L} = 2I_{slack} + \lambda = 0 \tag{6}$$

$$\nabla_X \mathcal{L} = \nabla_X F_{pf}^T(X_{pf})\lambda = 0 \tag{7}$$

$$F_{Ipf}(X_{pf}, I_{slack}, \lambda) = F_{Ipf}(X_{Ipf}) = \begin{bmatrix} \nabla_\lambda \mathcal{L} \\ \nabla_{I_{slack}} \mathcal{L} \\ \nabla_X \mathcal{L} \end{bmatrix} \tag{8}$$
$$= 0$$

### B. Power Grid Optimization

Power grid optimization applications, for example, AC-OPF, SC-OPF or transmission expansion planning, are represented as a constrained optimization (9) with non-linear AC network equality constraints ($g(V)$) and inequality constraints representing device limits ($h(V)$). The objective of the optimization ($f(V)$) varies depending on the application. To find a local minimum, most tools form a Lagrange equation (10) with Lagrange multipliers ($\lambda$) and complementary slack variables ($\mu$) [2]. A local minimum is solved for by a set of perturbed KKT equations (10)-(12), denoted $F_o$, shown in (14), that represent the necessary first order optimality conditions.

$$\min f(V) \\ s.t \tag{9}$$
$$g(V) = 0$$
$$h(V) < 0$$

$$\mathcal{L} = f(V) + \lambda^T g(V) + \mu^T h(V) \tag{10}$$

$$\nabla_V \mathcal{L} = \nabla_V f(V) + \nabla_V g^T(V)\lambda + \nabla_V h^T(V)\mu = 0 \tag{11}$$

$$\nabla_\lambda \mathcal{L} = g(V) = 0 \tag{12}$$

$$\mu^T h(V) + \epsilon = 0 \tag{13}$$

$$F_o(V, \lambda, \mu) = F_o(X_o) = \begin{bmatrix} \nabla_V \mathcal{L} \\ \nabla_\lambda \mathcal{L} \\ \mu^T h(V) + \epsilon \end{bmatrix} = 0 \quad (14)$$

For optimization, power flow and infeasibility power flow problems, we solve a set of non-linear equations representing either the steady state or the KKT equations that can be generalized to solving $F(X) = 0$, where $X$ is the general state vector based on the application. To solve the non-linear problem, the inherent solver uses Newton-Raphson (N-R).

N-R is especially powerful due to its quadratic convergence to the solution; however, it often displays slow convergence in the presence of bad initial conditions that are outside the basin of attraction. Bad initial conditions can cause stationary points, an infinite cycle, non-convexities leading to overshoot, or requires a damping factor to ensure convergence [19].

## III. NETWORK-STEPPING

In practice, initial conditions for both power flow and optimization problems are often derived from a previous known network that undergoes a discrete change, such as a topological change or a change in supply and demand. These initial conditions are often outside the basin of attraction for N-R and result in slow convergence. It is important to note that the initial conditions are, in practice, a feasible or optimal solution to a similar topological network. With this foresight, we introduce a successive relaxation method, known as a homotopy method, that gradually transforms the network from the previously solved network to the current one.

### A. Homotopy Methodology

A homotopy method is a mathematical successive relaxation technique that aims to solve a set of non-linear equations, $\mathcal{F}(X) = 0$, by embedding a scalar homotopy factor, $\gamma \in [0,1]$ to relax the problem to a trivial one represented by $\mathcal{G}(X) = 0$ [19]-[20]. A homotopy factor of $\gamma = 1$ relaxes the problem to a trivial problem and is iteratively decreased to a homotopy factor of 0 representing the original problem, $\mathcal{F}(X)$. The resulting series of problems, described by $\mathcal{H}(X, \gamma)$ (15) trace a path in the solution space from the trivial problem $\mathcal{G}(X)$ to the original, $\mathcal{F}(X)$. Each step of the homotopy factor presents a new problem, $\mathcal{H}(X, \gamma)$ that is solved with N-R. Fortunately, homotopy methods exploit the quadratic convergence of N-R to enable fast convergence between successive homotopy steps.

$$\mathcal{H}(X, \gamma) = (1 - \gamma)\mathcal{F}(X) + \gamma \mathcal{G}(X) = 0 \quad (15)$$
$$\text{where } \gamma \in [0,1]$$

### B. Network-Stepping Homotopy Method

We introduce a Network-Stepping homotopy method that defines the trivial problem, $\mathcal{G}(X)$ as the set of non-linear equations describing the previous known network, $\mathcal{N}^{k-1}$. The previous network has a set of non-linear equations, $F^{k-1}(X)$ describing either the steady state or the first order optimality conditions. The solution to the previous network, $X^{k-1}$, is the initial condition used to solve for the current network, $\mathcal{N}^k$ with corresponding non-linear equations, $F^k(X)$ representing the steady state or first order optimality conditions.

The Network-Stepping homotopy method embeds a homotopy factor in the non-linear equations to iteratively decrease the discrete change between the two networks, thereby transforming the previous network to the current one and tracing a path through the solution space, as depicted in Figure 1.

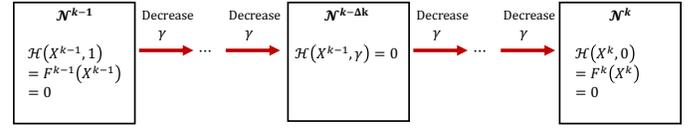

Figure 1 Network-Stepping Method gradually transforming from previous network to current

### C. Embedding Homotopy Factor as Current Injection

A homotopy factor, $\gamma$, is multiplied with the discrete change and maps a new set of sub-problems, $\mathcal{H}(X, \gamma)$ varying in the continuous domain of [0, 1]. To define the discrete change, consider the residual due to the solution of the previous network, $X^{k-1}$ in the set of non-linear equations for the current network, $F^k$. Denoted as $R$, the residual is a vector of mismatches due to the initial conditions (16).

$$F^k(X^{k-1}) = R \quad (16)$$
$$F^{k-1}(X^{k-1}) + R = R \quad (17)$$
$$F^k(X^{k-1}) = F^{k-1}(X^{k-1}) + R \quad (18)$$

It should be noted that the residual is in fact representing the discrete change between the non-linear equations of $F^{k-1}$ and $F^k$. The residual vector can be re-written as (17), knowing that the initial conditions, $X^{k-1}$ is a solution to the previous non-linear equations, $F^{k-1}$. It follows that we can re-write (16) as (18), thereby confirming that the residual describes the change between the networks.

In the case of power flow, $F^k = F^k_{pf}$, the residual vector represents a set of current injections at each bus equal to the current mismatch due to the discrete change. This is a direct consequence of the *substitution theorem* from circuit theory [19], that states any device model can be replaced by a current source equal to the current flowing through the device, without changing the overall circuit solution. For example, if the discrete change is that the current network, $\mathcal{N}^k$, opens a transmission line, the residual vector represents the current flowing through that transmission line in the previous configuration. In general, a discrete change in any device configuration between the two networks results in a residual vector representing the current injection equal to the change in the device current. Similarly, when considering optimization of the power grid, a discrete change between two networks will incur a residual vector, $R$ that represents the change in the primal set of equations (11) as well the dual (12)-(13).

After extracting the residual vector, the Network-Stepping methodology multiplies the homotopy factor by the residual to define a new set of problems given by (19).

$$\mathcal{H}(X, \gamma) = F^k(X) - \gamma R = 0 \quad (19)$$
$$\text{where } \gamma \in [0,1]$$
$$\mathcal{H}(X, 1) = F^k(X) - R = 0 \quad (20)$$
$$\mathcal{H}(X, 0) = F^k(X) = 0 \quad (21)$$

Evidently, the trivial problem, $\gamma = 1$, has the trivial solution of $\mathcal{H}(X^{k-1}, 1) = 0$, representing the solution to the set of non-linear equations for the previous network, $F^{k-1}$ (20). As the residual vector is iteratively reduced, we re-solve using the previous sub-problem as an initial condition for $\mathcal{H}(X,\lambda)$. During this process, the solution space is altered in successively shrinking contours to facilitate fast convergence to the next homotopy factor. Eventually, the homotopy factor reaches a value of 0, at which point, we arrive at the non-linear equations for the current network (21). By gradually reducing the support from the residual vector, we avoid oscillations and divergence generally seen in N-R in non-convex solution spaces. In fact, the solution trajectory is able to cut through the non-convex solution space, not only saving time but also achieving global convergence of the Network-Stepping homotopy method with slack current sources to indicate a feasible final solution.

### D. Global Convergence Proof

Global convergence for a general homotopy method is given by two requirements [20]:
i) The solution path created by the homotopy factor, $c(\gamma) = \mathcal{H}^{-1}(0)$ must be smooth and exist.
ii) If the path exists, it should intersect with the solution at $\gamma = 0$

The existence of the solution path is satisfied by including slack injections that act as feasibility sources. In fact, at any point during the homotopy process, it is uncertain if there exists a solution to $\mathcal{H}(X,\gamma) = 0$ due to an arbitrary residual vector. However, with the inclusion of feasibility sources, we can ensure there exists a feasible operation of $[X, I_{slack}]$ at each homotopy factor and converge at infeasible regions, as shown in Figure 3 [15]. Furthermore, the solution path requirement for smoothness and continuity is satisfied by the Implicit Function Theorem [15]. In using N-R to solve each sub-problem, this can be met by ensuring a full rank of the Jacobian $\mathcal{H}'(\gamma)$. With the existence and smoothness requirements met, we can safely ensure the first criteria of convergence is met.

The second criteria for existence of a solution at $\gamma = 0$ is also met through the use of slack injections [15] that ensures a solution regardless of the feasibility of the grid.

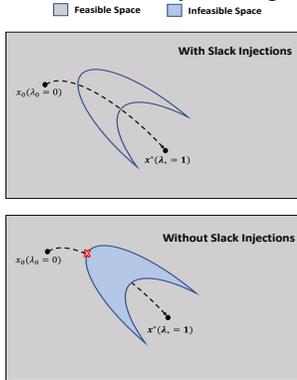

Figure 2 Homotopy path through infeasible region with and without slack injections [15].

With the criteria met, we ensure global convergence of the homotopy method. Importantly, the homotopy method is capable of passing through infeasible regions in the solution space shown in Figure 3, thereby avoiding excess iterations and damping steps and resulting in faster convergence.

## IV. APPLICATIONS AND RESULTS

The Network-Stepping methodology is generalizable for improving power flow analysis and optimization methods alike. It provides robust and timely convergence for many applications in both operation and planning settings. We highlight the capabilities and utility of Network-Stepping through synthetic real-grid examples including real-time AC-OPF, batch contingency analysis and stochastic power flow using the SUGAR [15] power flow simulation engine. The Network-Stepping approach is applicable to any current-injection based framework but requires the addition of slack current injection sources to ensure global convergence.

### A. Real-Time AC Optimal Power Flow

Operation engineers rely on a fast and accurate optimal power flow (OPF) to make real-time decisions in response to unexpected contingencies in small dispatch time windows (5-10 minutes). Given the solution state vector corresponding to the optimal network dispatch at the previous time window, we are able to use the Network-Stepping method to robustly and efficiently find the optimal dispatch of the current network. We demonstrate the scalability of the method on the 70k bus synthetic Eastern Interconnection [21], and compare it against fmincon in Matpower [22] using the same initial conditions that corresponds to an optimal base steady state solution. The network is modified to switch two transmission lines off and decrease the generation at renewables by 0.1% and loads in the corresponding area by 1%. The results show the convergence characteristics and time to reach a stationary point for three scenarios: i) Network-Stepping implemented in SUGAR (N-Stepping), ii) SUGAR Incremental Model Building (IMB) [15] and iii) Matpower optimal flow [22] as highlighted in Table 1. Results show that the Network-Stepping approach not only ensures convergence but also vastly improves the runtime.

Table 1 Real-Time AC-OPF convergence and Runtime Comparison between Network-Stepping methodology and Matpower fmincon

|  | N-Stepping | IMB | fmincon |
|---|---|---|---|
| Converged | YES | YES | NO |
| Total OPF cost | 3.10e6 | 3.10e6 | NA |
| Time (s) | 74 | 2198 | NA |
| Iterations | 23 | 572 | NA |

### B. Batch-Contingency Analysis

In practice, an optimally secure dispatch not only seeks the cheapest dispatch given a set of resources, but it also ensures feasibility across a set of contingencies. Future contingency analysis is likely to include the added feasibility analysis for a post-contingency steady state that also minimizes the cost of generation. This in fact becomes an OPF-like problem, as described by the ARPA-E Go competition [1], with added slack infeasibility sources, generator set-point constraints given from the base solution, and generator ramping limits. We denote the problem as a contingency OPF. Real-time operations require a fast and robust solution to abide by the small dispatch window. We demonstrate the utility of Network-Stepping by performing 1000 N-1 contingency analysis on the 70k bus synthetic Eastern Interconnection testcase [21]. The Network-Stepping method solves each contingency-OPF instance by a sequence of sub-problems described in Section III. The results of the batch contingency are compared against the robust Incremental

Model Building (IMB) implementation in SUGAR [15] and Matpower fmincon as shown in Table 2. Not only does the Network-Stepping implementation use fewer iterations due to the rapid homotopy method, but it also has a faster runtime due to the reduced post-processing. Future ISOs will likely parallelize the contingency-OPF on large on-premise cloud computers to improve runtime. Assuming access to 4 machines with 64 cores, the benefit of Network-Stepping becomes abundantly clear as we see nearly 7 times speed up in the total runtime shown in Table 2.

Table 2 Convergence and RunTime for Network-Stepping against IMB and fmincon for batch contingency-OPF on Synthetic Eastern Interconnection

|  | N-Stepping | IMB | fmincon |
|---|---|---|---|
| Convergence | 100% | 100% | 100% |
| Time (s) per Contingency | 79 | 2028 | 589 |
| Iterations per Contingency | 19 | 579 | 78 |
| Total Runtime (s) * | 308 | 7,922 | 2,301 |

*Parallelized runtime on 4 64-core machines

*C. Stochastic Power Flow*

Planning engineers are concerned with the stability of the system due to statistical variations for example from errors in weather prediction models or varying load conditions. Traditionally, a Monte-Carlo loop is used to determine the effect of statistical variations from load and resource patterns to the overall system. Often the constraining factor of a Monte-Carlo is the speed of each power flow simulation. The Network-Stepping method is useful for this application as it achieves fast convergence using the ideal scenario as the initial conditions. To demonstrate the efficacy, we perform a Monte-Carlo simulation on the 2000 bus Synthetic Texas grid [21] using the SUGAR Network-Stepping with 1000 samples taken from a Gaussian distribution of loads and renewable generation centered around the ideal case with a standard deviation of 20%. The voltage distribution for bus 1 is shown in Figure 3. The experiment serves to demonstrate the efficiency of solving the 1000 samples with an average runtime per sample of 12 seconds (23 iterations) and each sample remains feasible.

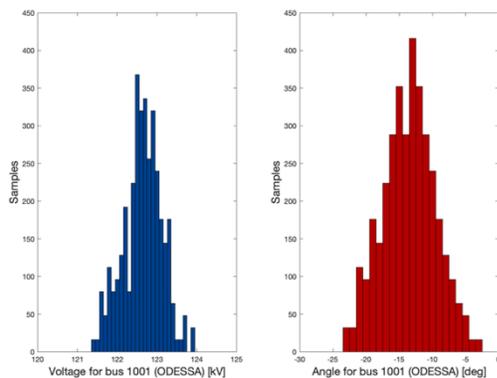

Figure 3 Stochastic results for voltage and angle of bus 1001,ODESSA, in Synthetic Texas testcase [21] with load and wind uncertainty

## V. CONCLUSION

With further reliance on power flow and optimization tools to operate and plan the grid, future engineers require a tool that is both fast and accurate. We introduce a novel homotopy method, Network-Stepping, that uses prior knowledge from a previous known state, to guide the solution trajectory. Network-Stepping guarantees a robust and fast convergence by gradually reducing the residual determined from the initial conditions. The proposed method is generalizable to power flow and optimization problems alike. In the results section we demonstrate the efficacy of the approach on few potential applications by improving the convergence robustness and simulation time of real-time OPF, batch contingency analysis and stochastic power flow with advantages seen when scaled to large systems such as the synthetic Eastern Interconnection.